\documentclass{article}

\usepackage{PRIMEarxiv}

\usepackage[utf8]{inputenc}	
\usepackage{hyperref}       	
\usepackage{url}            	
\usepackage{fancyhdr}    	
\usepackage{graphicx}      
\usepackage{amsmath}	
\usepackage[algo2e,ruled,vlined]{algorithm2e} 

\title{Functional imaging through scattering medium via fluorescence speckle demixing and localization}

\author{
  F. Soldevila$^1$, C. Moretti$^1$, T. Nöbauer$^2$, H. Sarafraz$^2$,\\ \bf{A. Vaziri}$^{2,3}$, and \bf{S. Gigan}$^{1,*}$ \\
  $^1$Laboratoire Kastler Brossel, ENS--Université PSL, CNRS, Sorbonne Université, College de France,\\ 24 Rue Lhomond, F-75005 Paris, France. \\
   $^2$Laboratory of Neurotechnology and Biophysics, The Rockefeller University, New York, NY, USA \\
   $^3$The Kavli Neural Systems Institute, The Rockefeller University, New York, NY, USA\\
  $^*$\texttt{sylvain.gigan@lkb.ens.fr}
}

\begin{document}
\maketitle

\begin{abstract}
Recently, fluorescence-based optical techniques have emerged as a powerful tool to probe information in the mammalian brain. However, tissue heterogeneities prevent clear imaging of deep neuron bodies due to light scattering. While several up-to-date approaches based on ballistic light allow to retrieve information at shallow depths inside the brain, non-invasive localization and functional imaging at depth still remains a challenge. It was recently shown that functional signals from time-varying fluorescent emitters located behind scattering samples could be retrieved by using a matrix factorization algorithm. Here we show that the seemingly information-less, low-contrast fluorescent speckle patterns recovered by the algorithm can be used to locate each individual emitter, even in the presence of background fluorescence. We test our approach by imaging the temporal activity of large groups of fluorescent sources behind different scattering phantoms mimicking biological tissues, and through a brain slice with a thickness of $\sim200$ $\mu m$.
\end{abstract}

\section{Introduction}
For the past decade, several light-based technologies have revolutionized the field of neuroscience\cite{boyden_millisecond-timescale_2005, deisseroth_optogenetics_2011, helmchen_imaging_2011}. In particular, genetically encoded calcium indicators (GECIs) have emerged as a powerful tool to monitor information processing in the brain in multiple animal models with high spatial resolution, contrast, and specificity \cite{chen_ultrasensitive_2013,helmchen_deep_2005,weisenburger_guide_2018,demas_high-speed_2021}. While these approaches allow to retrieve neuronal activity at high frame rates and even with subcellular resolution\cite{iyer_fast_2006, katona_fast_2012, prevedel_fast_2016}, several challenges in achieving large scale, deep, and possibly whole--brain neuronal activity recording in complex animal models such as mice are still present. In particular, tissue heterogeneities perturb the light wavefront as it travels through any sample, which limits the capability of any optical system to obtain sharp images (if the aberrations introduced by the medium are weak) or any spatial information at all (if the signal coming from the sample is fully scattered), at depth. In practice, this means that conventional microscopy is limited to imaging at depths that correspond to just a few mean free paths, which represents, at most, a few hundred microns in the brain\cite{ntziachristos_going_2010}. On the other hand, multi-photon fluorescence microscopy allows to penetrate deeper in a noninvasive manner, but it is still ultimately set by out-of-focus fluorescence\cite{weisenburger_guide_2018}. Alternatively, several micro-endoscopy approaches have emerged as a way to overcome this depth limit\cite{ohayon_minimally_2018, vasquez-lopez_subcellular_2018}, but have the drawback of being invasive.

Recently, wavefront shaping (WS) approaches have been proven to allow image retrieval through highly scattering media. The main idea of these techniques is to use a spatial light modulator (SLM) to introduce controlled changes on the wavefront, thus being able to compensate for the scattering events with either optimization procedures\cite{vellekoop_focusing_2007}, digital optical phase conjugation\cite{cui_implementation_2010}, or by measuring the transmission matrix (TM) of the system \cite{rotter_light_2017}. By doing so, it has been shown how it is possible to focus light through or inside scattering samples\cite{kim_transmission_2015, horstmeyer_guidestar-assisted_2015}, and using the so-called memory effect (ME), scan this focus in order to image small hidden objects\cite{freund_memory_1988, hsieh_imaging_2010, katz_looking_2012}. In fact, the use of the ME has also allowed to design systems that do not even need to use WS techniques to obtain useful information \cite{bertolotti_non-invasive_2012,katz_non-invasive_2014}. As of late, computation based approaches, merging ideas from both the optimization and the TM fields, have been shown to provide non-invasive fluorescence imaging inside scattering media even beyond the limitations established by the ME range\cite{boniface_non-invasive_2020, zhu_large_2022}. These approaches take advantage of the fact that the use of fluorescence contrast implies the incoherent addition of light signals coming from each individual emitter at the detector. When using time varying excitation, these signals can be efficiently unmixed by applying different matrix factorization approaches, such as non-negative matrix factorization (NMF). Taking advantage of this new paradigm, it has also been shown that NMF can be used to read out functional signals through the skull in mice from the incoherent, seemingly information-less fluorescent speckle patterns recorded from the sample \cite{moretti_readout_2020}.

Here, we tackle the problem of obtaining both the location and temporal activity (i.e. functional recording) of fluorescent functional signals through scattering media. From a set of fluctuating sources (mimicking a set of neurons), we show how it is possible to retrieve not only their temporal activity but also their spatial localization from the fluorescent signal reaching the detector. To do so, we rely on the fact that the signal coming from each source, after propagating through the scattering medium, generates a unique speckle pattern, which we refer to as fingerprint, at the sensor. All of these individual fingerprints are added incoherently at the image plane, where the sensor is placed. This generates a low-contrast image that will fluctuate over time according to the combined temporal activity of the emitters. From this fluctuating video, NMF can be used to decompose the recorded dataset into two different matrices, one containing the unmixed individual speckle fingerprints, and another one with their temporal activities.  Crucially, studying the correlations between the different fingerprints allows to retrieve the individual position of each emitter, thanks to the ME. This is carried out using a simple post-processing algorithm based on deconvolving each fingerprint from all overlapping fingerprints of other sources. We demonstrate the validity of the approach by retrieving the temporal activity and localizing large fluorescent sources behind scattering phantoms mimicking biological tissue and through a brain slice with a thickness of $\sim200$ $\mu m$. We also study the robustness of the process to background fluorescence. 

\section{Methods}
\subsection{Experimental Setup}

\begin{figure}[hbt!]
\centering
\includegraphics[width =0.95\linewidth]{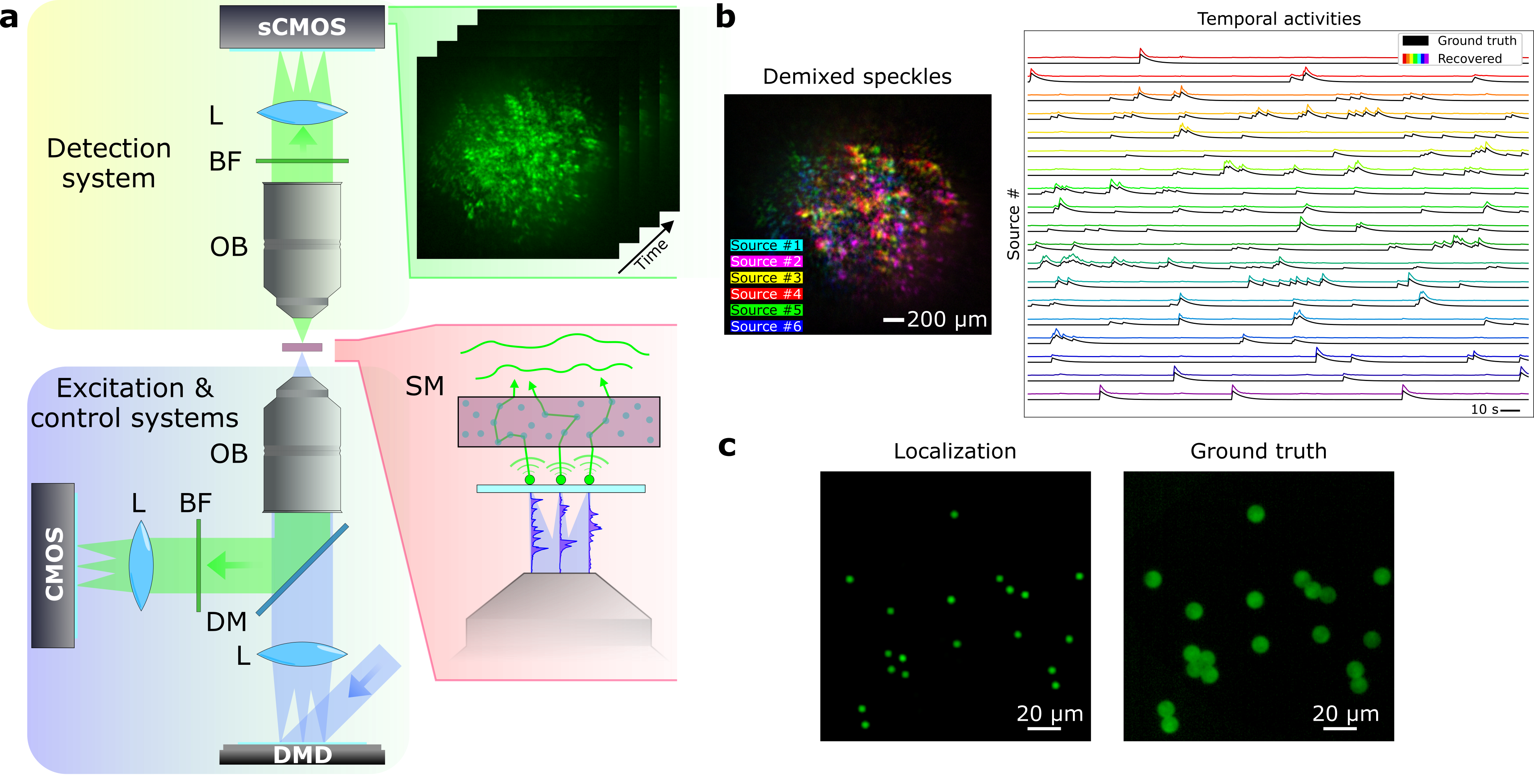}
\caption{ {\bf Experimental setup and principle of recovering the temporal activity and location of fluorescent emitters through scattering media.}  {\bf a}, A DMD is illuminated with a light source (blue laser) to excite a set of fluorescent beads with different spatio-temporal activations (red inset). The fluorescent signal propagates through the scattering sample, generating a set of speckle patterns that vary over time (green inset). The signals are collected by a microscope objective and a tube lens (L), and add incoherently on the camera (sCMOS). A band-pass filter (BF) removes any residual excitation light. A second imaging system records the ground truth spatial and temporal information (for control and comparison purposes). To do so, a dichroic mirror (DM) and a lens (L) are placed below the sample. {\bf b}, From the captured video, an unmixing algorithm allows to retrieve all the individual speckle patterns generated by each source and their temporal activities. {\bf c}, Studying the correlations between speckle patterns allows to retrieve the location of each individual emitter.}
\label{fig:fig1}
\end{figure}

The experimental system can be seen on Fig.\ref{fig:fig1}a. A 473 nm blue laser (LSR-0473-PFM-00100-01, Laserglow Technologies) was used to illuminate a digital micromirror device (DMD), the surface of which was imaged onto the sample plane by a tube lens (LA1708-A, Thorlabs) and the lower objective (Plan-NEOFLUAR $\times20$ 0.5 NA, Zeiss). We used the DMD (DLP LightCrafter 6500, Texas Instruments) to generate well-defined dynamic excitation patterns on a set of fluorescent sources, matching the dynamics of publicly available neuronal activity recordings\cite{chen_ultrasensitive_2013}. In our experiments, the samples consisted of groups of randomly distributed beads (FluoSpheres F8836, Thermofisher Scientific) with a diameter of 10 $\mu m$, similar in size to common neuron cell bodies\cite{bovetti_simultaneous_2017}. The samples extended over a field of view of about $160\times160$  $\mu m^2$, with bead densities ranging between $\sim47000$ beads/$mm^3$ and $\sim78000$ beads/$mm^3$, similar to the densities found in two-photon calcium imaging experiments in the mouse brain \cite{giovannucci_caiman_2019}. The emission spectrum of the beads is close to the common green fluorescent activity indicators. After excitation, the fluorescent signal propagated through a scattering medium and was imaged onto a scientific complementary metal--oxide--semiconductor (sCMOS) camera (Iris 15 sCMOS, Teledyne Photometrics) by a microscope objective (RMS10X PLAN ACHROMAT 0.25NA, Olympus) and a tube lens. A bandpass filter (MF530-43, Thorlabs) was used to block any excitation light from reaching the sensor. Additionally, we incorporated a control imaging system that directly imaged the sample plane in reflection, without going through the scattering medium. To this end, a dichroic beam splitter (DM, FF496-SDi01, Semrock) was used to collect the backpropagating fluorescent signal and the sample plane was imaged onto a CMOS camera (ACE2014-55um, Basler) by means of another tube lens. This allowed to obtain the ground-truth spatial position of all the emitters, and this information was only used a posteriori to verify the quality of our reconstruction.

After propagating through the scattering medium, the combination of the signals emerging from each emitter generated a time-varying, low-contrast speckle pattern that was recorded with the sCMOS camera. We performed background envelope removal by using high-pass filtering on each frame of the video in order to enhance contrast. After this, we fed the processed video to the NMF algorithm. This process generated two different matrices, one containing the individual spatial fingerprints from each emitter, and another containing the independent temporal activities. From the spatial fingerprints, we recovered the positions of each individual emitter by using a deconvolution approach\cite{zhu_large_2022}. We detail both procedures (unmixing and localization) in what follows.

\subsection{NMF-based unmixing}

Any recorded dataset can be expressed as a three-dimensional spatio-temporal object, $I(x,y,t)$. Given the fact that fingerprints from each emitter add incoherently onto the sensor, it is possible to write the $k^{th}$ frame of the recorded video as:

\begin{equation}
I_k(x,y) =\sum_{s=1}^{s=N}w_s(x,y) \cdot h_s(t),
\label{eq:eq1}
\end{equation}

where $k$ is the index of the frame (ranging from 0 to the number of frames in the video), $s$ enumerates each of the individual sources, and $h_s(t)$ and $w_s(x,y)$ correspond to the emission level and the individual speckle pattern generated by the $s^{th}$ source, respectively. In other words, any frame of the video can be expressed as a linear combination of a reduced number of fingerprints (corresponding to the $N$ sources in the sample), with their weights determined by their temporal activity. Then, it is possible to express the full dataset in matrix form as:

\begin{equation}
I = W \cdot H,
\label{eq:eq2}
\end{equation}

where each column of $I$ contains a reshaped frame of the video in vector form, the columns of $W$ contain all the individual fingerprints (reshaped in vector form), and the rows of $H$ encode the temporal activities of each source. The goal is to estimate both $W$ and $H$ from the observations, $I$. Expressing the system in matrix form allows to clearly see the whole retrieval procedure as a matrix factorization problem. Moreover, while the size of $I$ can be quite large (tens of thousands of pixels and hundreds of frames), the rank of the matrix is much smaller (equal to the number of sources, neglecting noise). This means that both $W$ and $H$ are much smaller than $I$. Due to the physical characteristics of the system (fluorescent signals, intensity measurements), both matrices can only have positive elements. This allows to take advantage of established low-rank non-negative matrix factorization frameworks, which tackle the inverse problem of retrieving both $W$ and $H$, given $I$, by solving the minimization problem:

\begin{equation}
\min_{W,H>0} \Arrowvert I -W\cdot H\Arrowvert^{2}_F.
\label{eq:eq3}
\end{equation}

To solve Eq.\ref{eq:eq3}, it is necessary to know the rank of $I$ (i.e., the number of sources). The rank can be estimated in a non invasive manner from the recorded dataset by comparing the outputs of different NMF runs with different ranks and minimizing the residual (see Supplementary Information). In practice, it is also possible to add regularization terms to the minimization problem that stabilize solutions and incorporate different priors. In the present case, it makes sense to add a sparsity constraint in the temporal domain (CEGI signals tend to have brief spikes, followed by longer decay times and periods of very low activity). Moreover, forcing some sparsity on $W$ helps retrieving higher contrast fingerprints, which makes localization easier. These regularization terms can be tuned depending on the experimental conditions (noise level, amount of a priori knowledge about the system, etc.) to improve the quality of the results and to reduce the post-processing time. In practice, we use the NMF solver contained in the {\it scikit-learn} package\cite{noauthor_sklearndecompositionnmf_nodate} (see Suplementary Information).

\subsection{Spatial localization}

Once the NMF procedure has been carried out, $W$ and $H$ provide the individual fingerprints and the temporal activities of each independent source, respectively. In previous work, the information contained in $H$ was shown to contain the functional signals from the sources\cite{moretti_readout_2020}. However, no information about the spatial position of the emitters was retrieved. Recently, it has also been shown in the context of structural imaging that it is possible to locate different emitters from their speckle fingerprints (even beyond the ME range) by using the information encoded in $W$\cite{zhu_large_2022}. The key idea is that neighboring emitters generate laterally shifted speckle patterns, and evaluating these lateral shifts reveals the $(x,y)$ positions of the sources. Moreover, even if the span of the source distribution goes beyond the ME range, the full location map can be retrieved under the condition that the sample is dense enough (i.e., if the maximum distance between any two neighboring sources is not longer than the ME range). In that case, it is possible to advance the position analysis from one emitter to its nearest neighbor by studying the correlation between their fingerprints, and build a full location map.

\begin{figure}[t!]
\centering
\includegraphics[width = 14cm]{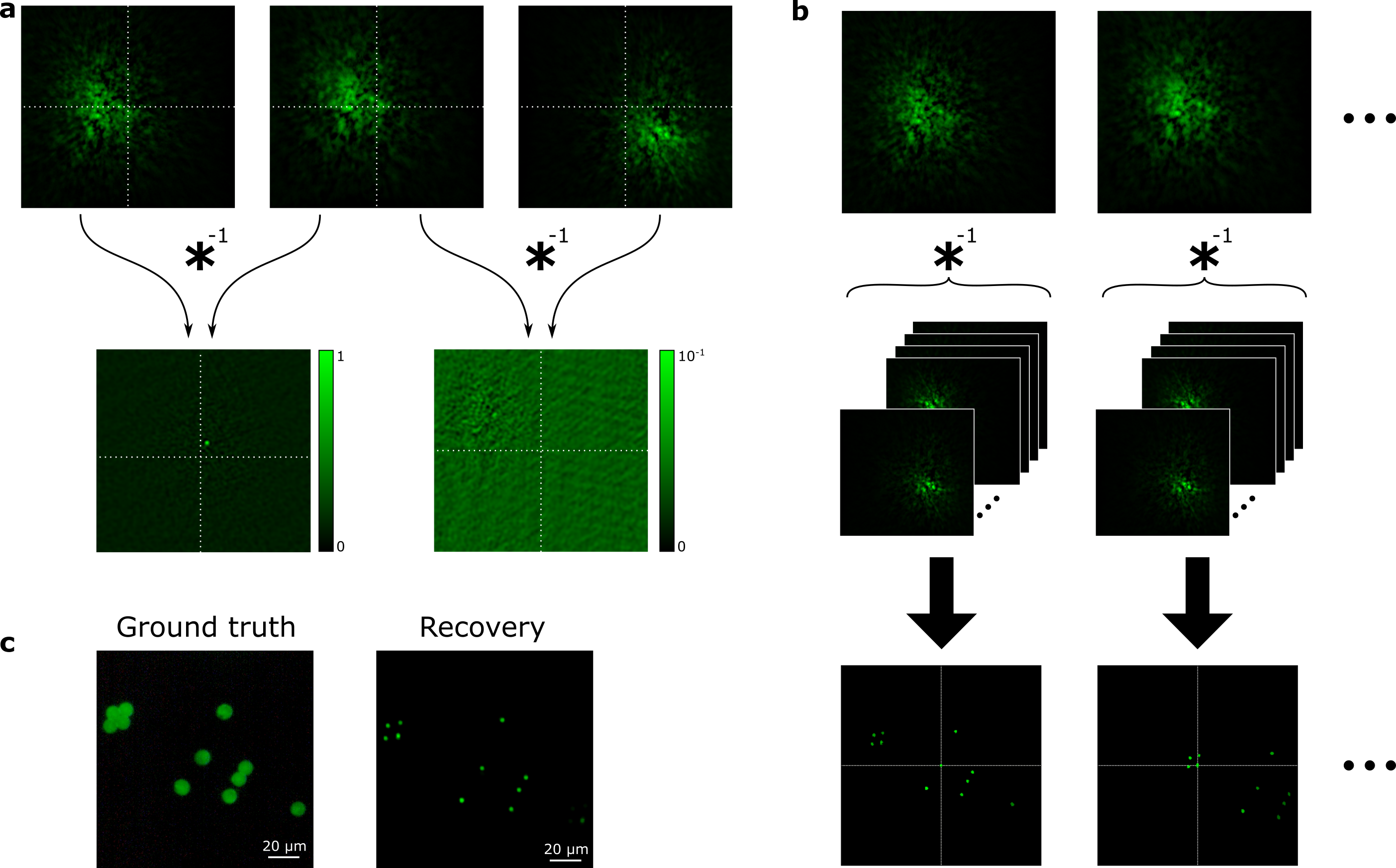}
\caption{{\bf Retrieving the spatial position from different fluorescent emitters by studying the correlations between their speckle fingerprints.} {\bf a,} Deconvolution between fingerprints coming from close (first and second) or far away (second and third) fingerprints. In  the first case, the two fingerprints are highly correlated, but laterally-shifted images, and the deconvolution yields a delta-like spike, which position from the center of the image provides the lateral shift between the two. In the second case, the two fingerprints come from emitters separated a distance longer than the ME range, and thus they are not correlated. In this scenario, the deconvolution provides a low-amplitude noisy image with no useful information. {\bf b,} For each emitter, deconvolution between its fingerprint and the fingerprints of other emitters provides a location map of emitters in its vicinity. {\bf c,} Stitching all the partial location maps provides the localization of all the emitters in the sample.}
\label{fig:fig2}
\end{figure}

Assuming perfect memory effect, the relationship between two fingerprints $w_i$ and $w_j$ can be written as the convolution of one of the fingerprints and a delta function:

\begin{equation}
w_i = w_j \ast \delta(x-x_0^{i,j},y-y_0^{i,j}),
\label{eq:eq4}
\end{equation}

where $x_0^{i,j}$ and $y_0^{i,j}$ account for the lateral shift between the two fingerprints ($i,j$).  Due to the ME, this shift is directly proportional to the relative position between the sources. Then, the lateral shift ($x_0^{i,j},y_0^{i,j}$) between any pair of fingerprints $w_i,w_j$, can be experimentally retrieved via a deconvolution ($\ast^{-1}$) between the two fingerprints:

\begin{equation}
\delta(x-x_0^{i,j},y-y_0^{i,j}) = w_i \ast^{-1} w_j.
\label{eq:eq5}
\end{equation}

The result of this deconvolution is a spike akin to a delta function, offset from the center by a distance that corresponds to the lateral shift between the two emitters (see Fig. \ref{fig:fig2}a). This is under the condition that the two fingerprints are indeed laterally shifted versions of each other (i.e., their respective sources lie within the ME range). In practice, due to the finite ME range, the deconvolved peak, which represents a correlation function, decreases in amplitude as the distance increases. For very distant emitters, the two fingerprints will not be correlated anymore, and the result of the deconvolution, rather than having that particular structure, will be a low amplitude noise--like pattern. Carrying out the deconvolution between all the possible pairs of fingerprints, and tracking the positions of the correlation peaks, allows to build a relative location map of all the emitters. For a particular emitter, $s$, it is possible to retrieve the partial location map ($M_s$) in its vicinity by adding the result of all the deconvolutions related to that emitter:

\begin{equation}
M_s = \sum_{i=1}^{i=N}w_s \ast^{-1} w_i.
\label{eq:eq6}
\end{equation}

This partial position map represents the relative positions, centered around the emitter $s$, of all the sources that lie at a distance from $s$ lower than the ME range (Fig. \ref{fig:fig2}b, lowest row of images). Note that this partial location map is retrieved by adding the result of all the deconvolutions (including the fingerprints that are not correlated). The correlated fingerprints will generate high amplitude spikes at the positions of the sources, and the uncorrelated fingerprints will yield a low--level noise background. By adding all the partial position maps (corrected by shifting all of them with respect to the same source) it is possible to obtain the full location map as:

\begin{equation}
M = \sum_{i=1}^{i=N}M_i(x-x_0^{1,i},y-y_0^{1,i}).
\label{eq:eq7}
\end{equation}

In our experiments, we performed this deconvolution by using a Wiener-Hunt approach from the {\it scikit-image} python module ({\it skimage.restoration.wiener})\cite{noauthor_module_nodate}. A full step-by-step analysis of the procedure can be found in Supplementary Information, and the code at \cite{noauthor_repository_nodate}.

\subsection{Results}

We tested our approach under different experimental conditions. First, we retrieved both the temporal activity and the spatial positions of a group of beads through a scattering medium consisting of a $\sim210$ $\mu m$ thick parafilm layer ($l_s \sim 170$ $\mu m$, $g \sim 0.8$ \cite{boniface_noninvasive_2019}). We designed our system with the possibility to vary the distance between the bead plane and the scattering material. This allowed us to slightly change the range of the ME while maintaining a condition of non-ballistic image information, and thus test the system's capability to retrieve extended objects. In a first experiment, the distance between the bead plane and the scattering phantom was set to 1.7 mm, corresponding to a very favorable situation with large ME range and signal-to-noise ratio (SNR). The source distribution consisted of 19 randomly distributed beads, which were excited with the DMD. A total recording of 500 frames with an integration time of 500 ms was acquired. An example of the images obtained under these conditions can be seen in the green inset of Fig.\ref{fig:fig1}. Fig.\ref{fig:fig1}b shows the ground truth temporal activities together with the ones retrieved by NMF and the fingerprints of a subset of six sources, superimposed in a false-color image, demonstrating the unmixing capability of the approach. It can be seen that some of these fingerprints (corresponding to close-by sources) are similar, i.e., exhibit high spatial correlation, but are shifted laterally. For example, sources number three and four (red and yellow) are related by a small horizontal shift. By using this information and the previously described procedure, it is possible to obtain an image of the relative location of the full group of nineteen sources, as shown in Fig.\ref{fig:fig1}c.

\begin{figure}[t!]
\centering
\includegraphics[width = 0.9\linewidth]{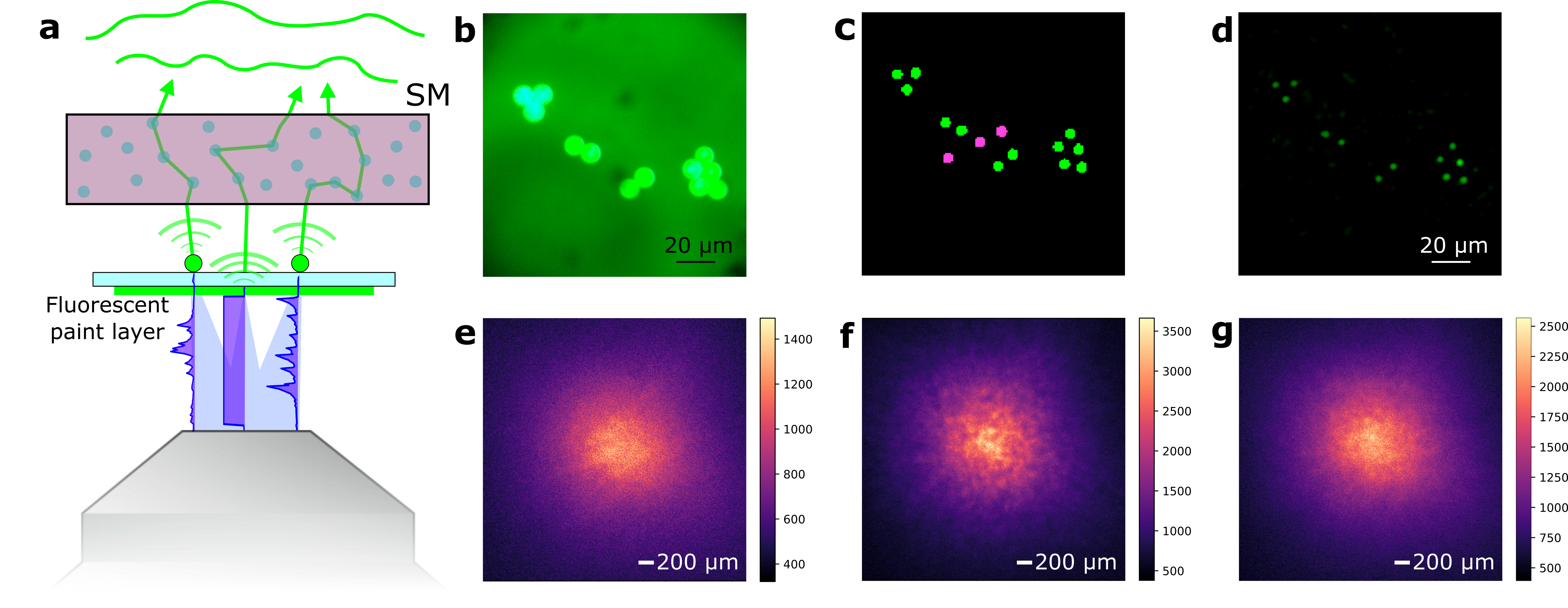}
\caption{ {\bf Retrieving emitter locations in the presence of background signal.} {\bf a}, Experimental configuration for the generation of out-of-focus fluorescent light. A thin layer of fluorescent paint is placed on the bottom side of the coverslip. As the excitation light passes through the sample, this layer generates fluorescent signal coming from a different axial plane from that of the sources. During the experiments, the DMD can be used to generate constant illumination of this layer of paint, thus generating a constant out-of-focus background that mimics auto-fluorescence signal commonly found in biological samples. {\bf b}, Image of the sample plane when using wide-field illumination. {\bf c}, Spatial mask generated on the DMD to excite both the sources (beads) and the background (paint layer). In this case, the paint layer is excited at three different spots (marked magenta) in the center of the field of view. {\bf d}, Retrieved localization of the sources after the unmixing procedure. {\bf e}, Speckle pattern captured by the detector when exciting only the fluorescent paint layer at the three spots shown in (c). {\bf f}, Speckle pattern generated onto the detector when exciting a single single source of the sample. {\bf g}, Single frame from the full recorded video, showing the low contrast speckle pattern that results form the combination of exciting both the fluorescent paint layer and a single emitter during the experiments.}
\label{fig:fig3}
\end{figure}

As a way to study more realistic scenarios for biological samples, we then moved to study the capability of the method to unmix and locate samples when the datasets are corrupted by background fluorescence, and with thicker volumetric scattering media. Under realistic conditions, the SNR of GECI signals is often very low due to scattering and attenuation in the sample. In order to mimic this condition, we built a sample consisting of two different depth planes, and increased the thickness of the scattering phantom. We placed beads on one side of a microscope coverslip and a thin uniform layer of fluorescent paint on the other. With such a sample, our illumination system allows us to excite both the beads (to mimic neurons) and the paint layer (to mimic out-of-focus background fluorescence). Thus, we could excite a region with an arbitrary size during the full dataset recording, which reduced the contrast of the recorded dataset in a controlled manner. We show a schematic of this system in Fig.\ref{fig:fig3}a. Under brightfield illumination, the beads appear as bright disks on a diffuse background (Fig.\ref{fig:fig3}b). In our experiments, we tested different background levels to determine the highest amount of background that the system could tolerate before failing to unmix and locate all the sources. To generate the background, we excited a region of the sample containing only fluorescent paint and no beads with a constant signal. An example of the spatial mask generated on the DMD to illuminate the sample can be seen in Fig.\ref{fig:fig3}c. Here, the green disks correspond to the regions of the DMD that are imaged onto the beads, and the magenta disks generate the constant background signal from the out-of-focus plane.  Fig.\ref{fig:fig3}d shows the locations of the sources present in the sample. In this case, the scattering sample consisted of several layers of parafilm with a total thickness of 0.75 mm, placed at a distance from the beads of 0.65 mm. To visualize the loss in contrast due to the background, Fig.\ref{fig:fig3}e-g show different examples of the speckle patterns recorded by the camera when only the background is excited (e), when a single source is excited (f), or when both the background and a single source are excited (g). Also, it is possible to estimate the signal-to-background ratio (SBR) of our measurements by calculating the total energy of the speckle patterns associated with the background or each one of the individual sources. By doing so, we found out that the system was able to unmix and locate all the sources down to an SBR of approximately $1.6$.

\begin{figure}[hbt!]
\centering
\includegraphics[width = 8cm]{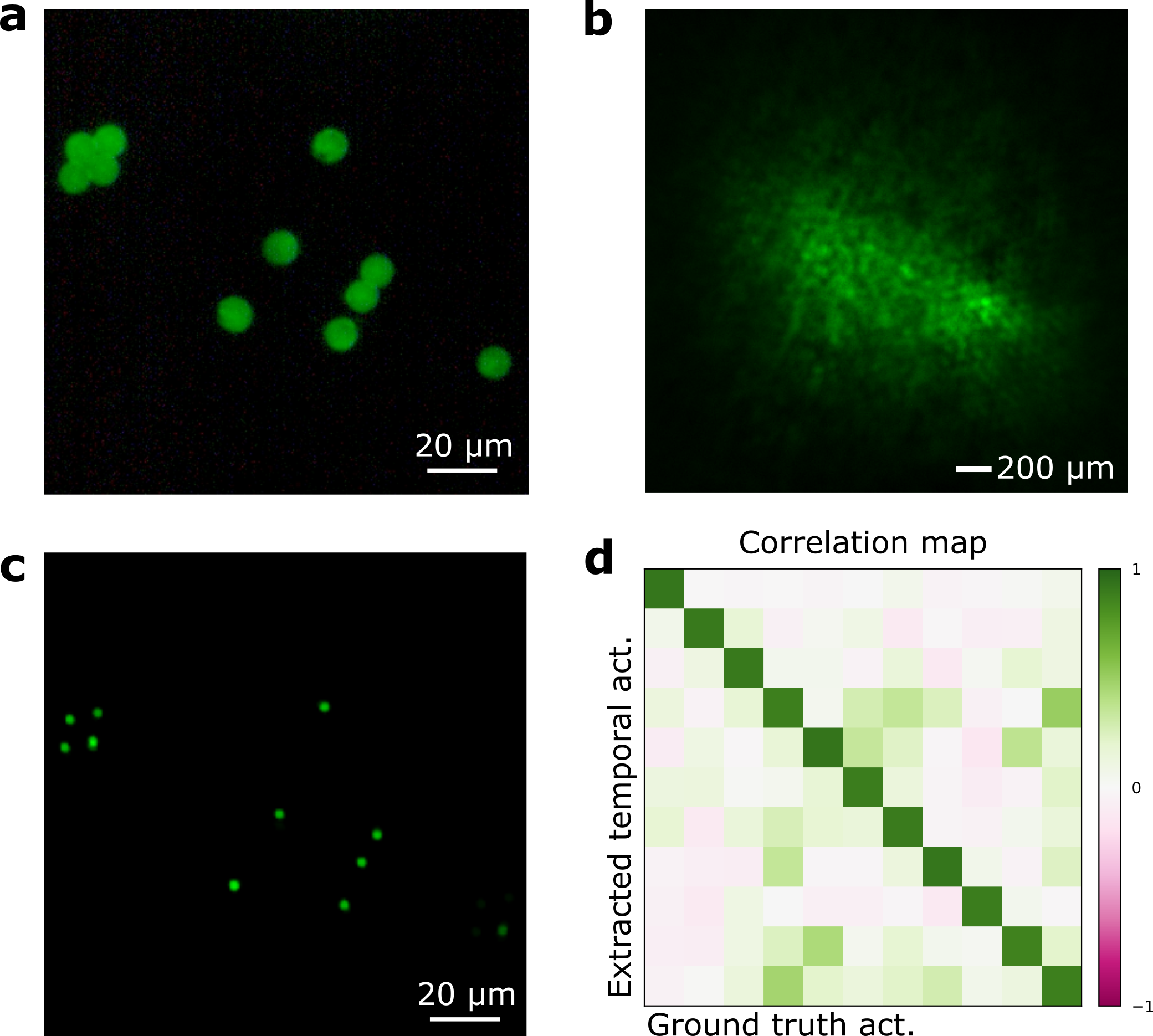}
\caption{ {\bf Localization and temporal activity retrieval through biological tissue.}  {\bf a}, Ground truth source distribution. {\bf b}, Bright-field image through a 200 $\mu m$ brain slice. {\bf c}, Localization retrieval of the sources in the field of view. {\bf d}, Temporal activity comparison between the retrieved traces and the ground truth excitations on the sources.}
\label{fig:fig4}
\end{figure}

Last, we tested our system with biological tissue as the scattering medium. In this case, we placed a 200 $\mu m$ fixed brain slice at a distance from the beads of $\sim 680$ $\mu m$, and recorded a full dataset (500 frames with an integration time of 2 s). Fig.\ref{fig:fig4}a shows the ground truth bead distribution, with a total extent of about 130 $\mu m$. Under bright-field illumination, the image retrieved by the system was a low--contrast speckle pattern, as shown in Fig.\ref{fig:fig4}b, which does not contain any obvious spatial information about the source location. After the NMF procedure, we were able to retrieve the spatial positions of all the sources (Fig.\ref{fig:fig4}c). In Fig.\ref{fig:fig4}d we show a comparison between the retrieved temporal traces and the ground truth activities that we used to excite each source. To do so, we calculated the correlation coefficient between all the pairs of traces, showing a very good agreement. The range of the ME was estimated to be about 75 $\mu m$, so the full source distribution extended approximately 1.7 times the ME range.

\section{Discussion}

In this work, we have shown that it is possible to retrieve not only the individual temporal traces, but also the location of fluorescent, temporally modulated, extended sources through scattering media. We have demonstrated the principle using both phantoms and fixed biological tissue as scattering samples, with a number of sources in the order of a few tens, randomly distributed in a 2D plane, and with temporal activities extracted from public available neuronal recordings. Even though the scattering introduced by tissue strongly scatters the light distribution emanating from the sources, an unmixing algorithm based on NMF allows to retrieve the individual speckle fingerprints coming from each emitter without requiring any ballistic information. Studying the correlation between different fingerprints provides the location of each individual source, even beyond the ME range. 

While the proposed technique is able to provide useful information in this proof-of principle but realistic scenario, some challenges still hinder its application {\it in vivo}. First, there are many applications where it is of interest to retrieve the temporal activities from emitters placed at multiple depth planes. Although the NMF algorithm is not affected by this \cite{moretti_readout_2020}, sources located at different axial positions will generate fingerprints that will not be related by just a lateral shift. In order to retrieve the location in these cases, more sophisticated analyses, based on lateral shifts and scale changes, should be explored \cite{okamoto_noninvasive_2019}. Also, while the method allows to retrieve the location of sources even if they span over large distances, close neighboring emitters are still required to lay inside the ME range. Even though this is a strong requirement, neural networks present in the brain show intricate and highly-packed neuron distributions, which should help fulfilling this prerequisite. Second, precise localization depends on the NMF algorithm successfully unmixing the fingerprints. For this to happen, the recorded frames need to present high enough contrast values. While multiple parameters reduce the contrast during the experiments, the most relevant are the number and size of sources present in the sample and out-of-focus fluorescence. Given the nature of the emitters, fingerprints are added incoherently at the sensor, generating a speckle pattern which contrast decreases with the number of sources as $1/\sqrt{N}$. On top of that, out of focus fluorescence will further decrease this value by introducing a spurious signal which does not contain any information about the temporal activity of the sources. In order to extend the number of sources from tens to hundreds, we will probably need to add more priors to the NMF algorithm with the aim to compensate for the drop in performance due to the contrast loss could be studied. Last, using different contrast mechanisms, such as multiphoton fluorescence, would help reducing out-of-focus signal, increasing the SBR to levels where successful unmixing can be achieved.


\section*{Acknowledgments}
Research reported in this publication was supported by the National Institute of Neurological Disorders and Stroke of the National Institutes of Health under award number 1RF1NS113251 (A.V.) and the Kavli Foundation through the Kavli Neural System Institute (A.V.), in particular through a Kavli Neural Systems Institute postdoctoral fellowship (T.N.). We thank L. Bourdieu, W. Akermann, and F. Xia for providing biological samples.

\section*{Disclosures} The authors declare no conflicts of interest.

\clearpage

\bibliography{biblio}

\clearpage

\section*{Supplementary Information}

\subsection*{NMF rank estimation}

\begin{figure}[hbt!]
\centering
\includegraphics[width = 12 cm]{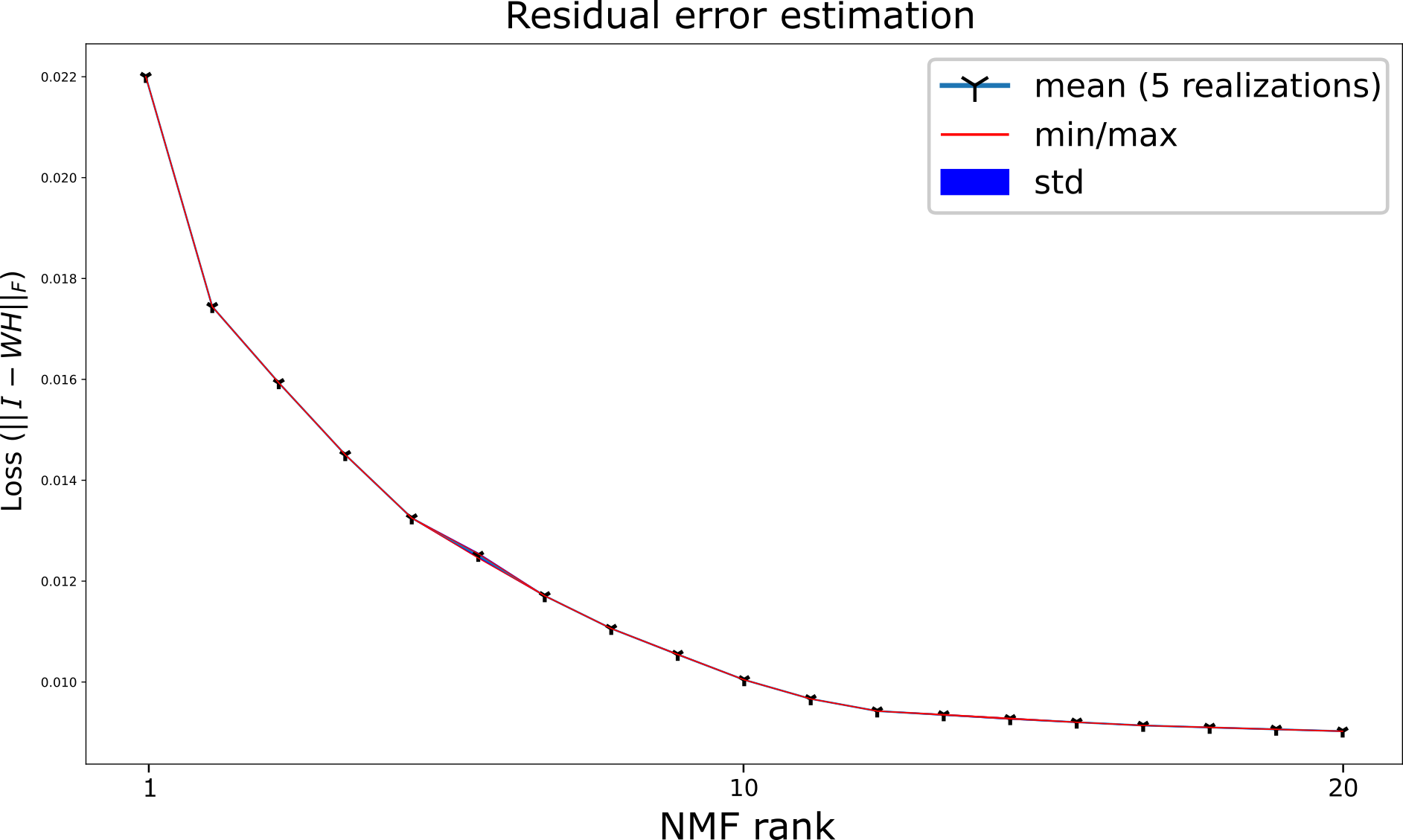}
\caption{{\bf Rank estimation from experimental data.}  For the same recorded dataset, we show the average residual error for five different NMF realizations (with different random initializations) for different rank values. After the rank surpasses the true number of sources in the sample, the residual error decreases at a much lower rate, a phenomena that can be used to estimate the number of emitters in the sample.}
\label{fig:figS1}
\end{figure}

In order to solve the NMF minimization procedure, the rank of the system needs to be set. This rank is, in an ideal scenario (without noise), the number of sources present in the sample. Given the fact that this number is in principle unknown, a method to estimate it from the experimental data was used. When factorizing the input matrix, $I$, it is possible to set the rank, $k$, to any value between 1 and the maximum possible rank of the dataset. Then, it is possible to check the quality of the factorization by looking at the residual error from the NMF ($\Arrowvert I -W_{est}\cdot H_{est}\Arrowvert^{2}_F$). When the rank of the system is underestimated, the factorization procedure tends to merge multiple fingerprints (temporal activities) into single columns (rows) of $W_{est}$ ($H_{est}$). This implies a high residual error, which becomes smaller as the rank approaches the correct number of sources in the sample. As the rank gets higher than the number of sources, the factorization starts using columns (rows) of $W_{est}$ ($H_{est}$) to fit the noise present in the measurements, which further reduces the residual error, but at an almost linear rate and with a very small slope. We use this change in the reduction rate of the residual error to estimate the number of sources of the sample. In Fig.\ref{fig:figS1}, we show the estimation for the brain slice dataset in the main text with ranks ranging from 1 to 20. In this case, the true number of emitters was 11, which is in good agreement with the region where the slope of the curve changes. Although this method is not exact, we experimentally find that a slight overestimation of the rank neither hinders the capability of the system to retrieve the temporal activities nor the spatial position of the sources. The extra fingerprints recovered when overestimating the rank tend to be high contrast noise-like images, which are discarded by the deconvolution-based localization procedure. In the same manner, the temporal activities from the extra rows of H present random-like signals clearly different from neuronal activity.

\subsection*{NMF inversion problem}

In order to solve the general NMF problem, multiple numerical methods can be used\cite{berry_algorithms_2007,ho_descent_2011}. While many of the currently available solvers simply tackle the simplest form of the inversion problem found in the main text ($\min \Arrowvert I -W\cdot H\Arrowvert^{2}_F$ subject to $W, H > 0$), it is possible to add regularization terms with some a priori information about the system, such as the sparsity of either the fingerprints or the temporal activities of the sources in the sample. Thus, we can formulate the NMF problem as:

\begin{align*}
&\min_{W,H>0} 0.5 \cdot ||I - WH||_{\beta}^2 \\
&+ \alpha_W \cdot l_{1r} \cdot n_{pixels} \cdot ||vec(W)||_1 \\
&+ \alpha_H \cdot l_{1r} \cdot n_{frames} \cdot ||vec(H)||_1 \\
&+ 0.5 \cdot \alpha_W \cdot (1 - l_{1r}) \cdot n_{pixels} \cdot ||W||_{F}^2\\
&+ 0.5 \cdot \alpha_H \cdot (1 - l_{1r}) \cdot n_{frames} \cdot ||H||_{F}^2,\end{align*}

where $||A||_{F}^2 = \sum_{i,j} A_{ij}^2$ corresponds to the Frobenius norm of a matrix, $A$, $||vec(A)||_1 = \sum_{i,j} abs(A_{ij})$ corresponds to the $l_1$ element-wise norm, and $||I - WH||_{\beta}$ represents the desired $\beta$-norm to calculate (1, 2, $\le 0$). Here, the additional terms introduced account for both the sparsity of $W$ and $H$, and are governed by $\alpha_W$, $\alpha_H$, and $l_{1r}$ (with two scaling factors, $n_{pixels}$ and $n_{frames}$, accounting for the vast differences between the number of elements of $W$ and $H$). Both $\alpha_W$ and $\alpha_H$ can take different values in order to weight the strength of the regularization between $W$ and $H$, and $l_{1r}$ can be used to continuously choose between different penalty forms. For the limit $l_{1r} = 0$, the penalty behaves like a standard Frobenius norm, while $l_{1r} = 1$ corresponds to an element-wise $l_{1}$ penalty (favoring sparsity). In our case, we acquire low contrast images that result of  the incoherent addition of many highly-contrasted individual speckles. Furthermore, these patterns do not fully cover the field-of-view of the camera, so some degree of sparsity is to be expected on each individual fingerprint. Moreover, this promotes recoveries where the fingerprints have higher contrast, which greatly helps the localization procedure. Last, the temporal activities that we use to mimic neuronal activity consist of short bursts of activity, usually followed by longer decay times and periods of little activity, so it is reasonable to consider some sparsity on the recovery of $H$. We find that, in our experimental conditions, a good compromise between fidelity, sparsity, and recovery time is found with $\beta = 2$, $\alpha_W = 1.5$, $\alpha_H = 0.5$, and $l_{1r} = 0.5$. While these regularization parameters have to be manually tuned and are experiment-dependent, several approaches to automatically estimate their values could be explored in the future\cite{mead_newton_2009, bardsley_regularization_2009}. The full system is solved by using the {\it scikit-learn} NMF package\cite{noauthor_sklearndecompositionnmf_nodate}, and it takes a few minutes to compute for datasets consisting of 500 frames with resolutions in the order of $300 \times 300$ pixels using a desktop CPU (Intel i7-9700) with 16 Gb of RAM. Bigger datasets and/or faster reconstruction times could be reached by using GPU-based implementations, but this lies outside of the scope of this work.

\subsection*{Step-by-step localization procedure}
Here, we introduce the post-processing workflow to obtain the location of the emitters from the recorded dataset. First, we crop and filter the frames recorded by the camera (as the sensor is larger than the area covered by the speckle patterns). Then, we remove the intensity envelope by high-pass filtering. In the experiments where there is a constant signal present (as in Fig.\ref{fig:fig3}), we perform a rank--1 NMF to identify this constant component in the dataset, which we later use to initialize both $W$ and $H$ when performing a full--rank NMF with the rank set to the estimated number of emitters. This helps unmix the time--varying fingerprints and the background present in all the frames due to the constant fluorescence signal. Otherwise, we just initialise the NMF with the Nonnegative Singular Value Decomposition (NNSVD) of the recorded dataset. After the NMF is performed, we deconvolve all the fingerprints in pairs to locate the shifts between them, and finally we merge all the information in the full location map. The codes can be found at \cite{noauthor_repository_nodate}.

\begin{algorithm2e}
\DontPrintSemicolon
\SetAlgoLined
\KwResult{Returns the location map of the emitters in the sample by performing NMF over the recorded dataset. The fingerprints provided by the NMF are deconvolved to calculate the distances between the different sources.}\;

Post-process recorded dataset: select the region of the sensor with speckle patterns (cropping), perform binning (reduce size to increase speed) and high-pass filtering (remove envelope, increase contrast)\;
\;
\If{constant background = True}{
	Do Rank-1 NMF to estimate constant background in the dataset\;
}
\;
Perform  a full--rank NMF on the recorded dataset ($n_s$ = number of sources)\;
\eIf{constant background = True}{
	Set $rank = n_s + 1$\;
	Perform the NMF, initializing $W$ and $H$ with the result of the rank-1 NMF\;
	}{
	Set $rank = n_s$\;
	Perform the NMF, initializing $W$ and $H$ with the Nonnegative Singular Value Decomposition (NNSVD) of the recorded dataset\;
}
\;
Calculate source positions by deconvolving the fingerprints provided by the NMF algorithm.\;
\For{$i=1:n_s$}{
	\For{$j=1:n_s$}{
		Deconvolve $fingerprint_i$ and $fingerprint_j$\; $\delta(x-x_0^{i,j},y-y_0^{i,j}) = w_i \ast^{-1} w_j$\;
		Calculate distance between $source_i$ and $source_j$ as the position of the delta--like peak $(x_0^{i,j},y_0^{i,j})$\;
	}
	Combine deconvolutions in a partial location map ($M_i$):\;
	$M_i = \sum_{j=1}^{j=n_s}w_i \ast^{-1} w_j$\;
}\;
Correct shifts between all the partial location maps by using the distances between the sources, then add together to generate the full location map\;
$M = \sum_{i=1}^{i=n_s}M_i(x-x_0^{1,i},y-y_0^{1,i})$

\caption{Step-by-step localization procedure}
\label{alg:a1}
\end{algorithm2e}

\end{document}